\documentclass[12pt]{article}

\usepackage{amssymb}

\usepackage{hyperref} 

\usepackage[autostyle=true,german=quotes]{csquotes} 

\usepackage{amsmath,amssymb,amsthm,amsfonts,amsbsy,latexsym}

\usepackage{caption}
\usepackage{subcaption}

\usepackage{longtable}

\usepackage{graphicx}

\usepackage{wrapfig}
\usepackage{graphicx} 
\usepackage{placeins}
\usepackage{tikz}
\usetikzlibrary{shapes.geometric, arrows}
\tikzstyle{startstop} = [rectangle, rounded corners, minimum width=3cm, minimum height=1cm,text centered, draw=black, fill=blue!30]
\tikzstyle{io} = [trapezium, trapezium left angle=70, trapezium right angle=110, minimum width=3cm, minimum height=1cm, text centered, draw=black, fill=blue!30]
\tikzstyle{arrow} = [thick,->,>=stealth]
\usepackage{tabulary}
\usepackage{tabularx}

\theoremstyle{definition}

\usepackage{algorithm}
\usepackage{enumerate}
\usepackage{algpseudocode}
\usepackage{bm} 
\usepackage{makecell}

\usepackage{lmodern}
\usepackage{dsfont}
\usepackage{bbm}
\usepackage[round]{natbib}
\usepackage[defaultlines=2,all]{nowidow}
\usepackage{caption}
\usepackage[labelformat=simple]{subcaption}
\usepackage{ragged2e}
\usepackage{threeparttable} 
\usepackage{placeins}

\usepackage{multirow}
\usepackage{array}
\usepackage{hhline}
\usepackage{xcolor}
\usepackage{booktabs}
\usepackage{longtable}
\usepackage{siunitx} 
\usepackage{csvsimple}

\usepackage[normalem]{ulem}
\usepackage{listings}

\definecolor{codegreen}{rgb}{0,0.6,0}
\definecolor{codegray}{rgb}{0.5,0.5,0.5}
\definecolor{codepurple}{rgb}{0.58,0,0.82}
\definecolor{backcolour}{rgb}{0.95,0.95,0.92}

\lstdefinestyle{mystyle}{,   
    commentstyle=\color{codegreen},
    keywordstyle=\color{codegreen},
    numberstyle=\tiny\color{codegray},
    stringstyle=\color{codepurple},
    basicstyle=\ttfamily\scriptsize,
    breakatwhitespace=false,         
    breaklines=true,                 
    captionpos=b,                    
    keepspaces=true,                 
    numbers=left,                    
    numbersep=4pt,                  
    showspaces=false,                
    showstringspaces=false,
    showtabs=false,                  
    tabsize=2
}

\usepackage[nameinlink,capitalise,german]{cleveref}

\newcommand{\mytilde}{{\raise.17ex\hbox{$\scriptstyle\mathtt{\sim}$}}\xspace}

\usepackage{mathtools}

\usepackage{color}

\usepackage[nottoc]{tocbibind}

\usepackage{natbib}
\usepackage{hyperref}                   
\hypersetup{                            
    colorlinks=true,                    
    linkcolor=black,                     
    citecolor=black,                     
    urlcolor=black                       
}

\title{ A Machine Learning-based Anomaly Detection Framework in Life Insurance Contracts}

\author{Andreas Groll\thanks{Department of Statistics, TU Dortmund University},  Akshat Khanna\footnotemark[1], Leonid Zeldin\thanks{Department of Statistics, TU Dortmund University; corresponding author: leonid.zeldin@tu-dortmund.de}}
\date {November 2024}

\begin{document}

\maketitle

\mbox{}
\vfill

\pagebreak

\begin{abstract}



\noindent 

\noindent Life insurance, like other forms of insurance, relies heavily on large volumes of data. The business model is based on an exchange where companies receive payments in return for the promise to provide coverage in case of an accident. Thus, trust in the integrity of the data stored in databases is crucial. One method to ensure data reliability is the automatic detection of anomalies. While this approach is highly useful, it is also challenging due to the scarcity of labeled data that distinguish between normal and anomalous contracts or inter\-actions. This manuscript discusses several classical and modern unsupervised anomaly detection methods and compares their performance across two different datasets. In order to facilitate the adoption of these methods by companies, this work also explores ways to automate the process, making it accessible even to non-data scientists.


\end{abstract}


\section{Introduction}
In the life insurance industry, where promises of financial protection are intertwined with complex agreements, ensuring smooth operations is critical. The foundation of insurance companies is built on vast databases containing detailed information on policies, premium payments, and beneficiaries. Any unusual pattern, transaction, or payment within this data can pose a significant risk to the company, its customers, and the overall trust in the insurance sector. These irregularities, known as anomalies, could signal data errors, fraud, or emerging trends that require immediate attention.

The process of identifying such unusual patterns that deviate from expected norms is called anomaly detection. In this context, the terms ``anomalies" and ``outliers" are often used interchangeably. Detecting anomalies is crucial as it can reveal actionable insights, especially in fields such as healthcare or security (see e.g. \citet{AnomalyDetection}), and has been studied in the statistical community since as early as the 19th century (see \cite{Edgeworth1887}).

Typically, such anomalies are unpredictable and cannot be foreseen, making it challenging to train a model on labeled data since such data is often nonexistent or artificially created, far removed from real-world issues. In this case, unsupervised techniques are required, which attempt to identify irregularities in the data without prior knowledge of their exact nature.

This work examines several classical proximity-based methods, along with a tree-based method applied in an unsupervised manner, to analyze two different datasets. These analyses will be compared with the results produced by modern deep learning techniques, specifically autoencoders and variational autoencoders. To evaluate the performance of these models, we introduced four manually created anomalies into each dataset, simulating anomalous behavior at the contract level rather than focusing solely on unusual variable values. The datasets used are derived from open-source databases, as real-world data for such analyses is not publicly available. The two most suitable open-source datasets are from the health insurance sector, which closely resembles life insurance data in terms of structure and content.

The remainder of this manuscript is structured as follows. Section 2 provides the motivation for this work and positions it into the existing literature. In Section 3, we introduce the aforementioned datasets, discussing their size, available variables, and suitability for the intended application. The core of the work is found in Section 4, which discusses the methods employed in this study. This section covers proximity-based methods, including the Silhouette Score for clustering evaluation, as well as the tree-based method Isolation Forest, which is evaluated using the Anomaly Score. Additionally, deep learning techniques such as autoencoders and variational autoencoders are explored, with their performance assessed through the Reconstruction Error. Section 5 begins with an introduction to data preprocessing techniques, including one-hot encoding for handling categorical variables, which are then applied to the two chosen datasets. This is followed by a discussion of the results when applying the methods presented before. Finally, Section 6 summarizes the findings, emphasizing how deep learning methods outperformed basic machine learning approaches. The section concludes with recommendations for future research and development.

\section{Motivation}

The primary focus of this work is to address the challenges and key difficulties associated with unsupervised anomaly detection. Life insurance involves a delicate structure where every step is meticulously overseen by numerous stakeholders. Their goal is to ensure that both the insurer and the insured are well-protected, and that the integrity of operations is maintained. Anomalies detected in life insurance contracts can signal potential fraud, data errors, or unidentified patterns that require thorough investigation and immediate human intervention. If such anomalies in the database are overlooked, it could lead to significant financial losses for the company and ultimately erode public trust in the insurance process and the companies involved. To begin, it is crucial to define what we mean by anomalies.

\begin{figure}[!htb]
   \centering
\includegraphics[width=0.7\textwidth]{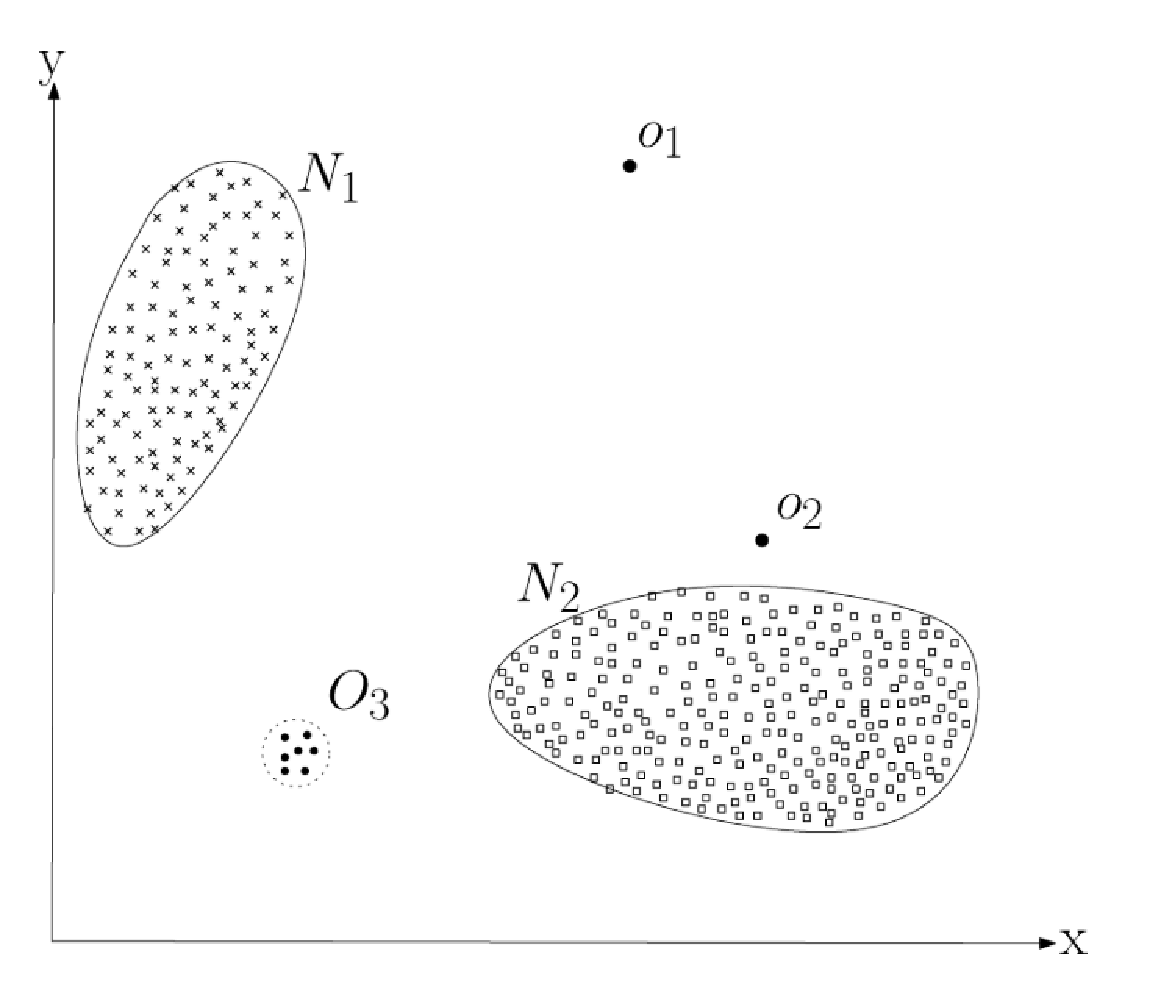}
    \caption{An example representation of anomalous and normal points in an example data set with two variables, $X$ and $Y$, taken from \citet{chandola2009survey}.}
    \label{fig:Anomalies}
\end{figure}

In general, anomalies, or outliers, are patterns in data that deviate substantially from what is expected as typical behavior. An example of anomalies in a basic two-dimensional dataset is illustrated in Figure~\ref{fig:Anomalies}. Most observations in this dataset cluster within two primary normal zones, labeled $N_1$ and $N_2$, and are considered normal points. In contrast, the points located within region $O_3$ and those that are significantly distant from these regions, such as points $o_1$ and $o_2$, are identified as anomalies. This classification arises because, unlike the dense regions $N_1$ and $N_2$, these points are situated in areas of very low density and are located far from the typical data distribution.

Traditionally, anomaly detection methods are supervised and rely on labeled datasets where instances of normal and anomalous behavior are pre-identified. Prior work in the fields of anomaly detection proposed that supervised anomaly detection methods generally outperform unsupervised or hybrid methods (see e.g. \citet{chandola2009survey, bauder2017medicare}). Amongst all the supervised methods, support vector machines (SVM, \cite{hu2003robust}) and neural networks (see e.g. \citet{callegari2014neural, chalapathy2018oneclass}) are the ones that have shown impressive results on various labeled benchmarks. However, typical life insurance datasets are vast and high-dimensional, making it impractical to obtain datasets with pre-labeled anomalies. Labeling anomalies is a labor-intensive, costly process for companies, and often impossible due to the sheer volume of data. This is the primary reason why exploring unsupervised machine learning methods, which can identify anomalies without pre-labeled datasets and with minimal human intervention, is essential, as described in~\citet{sun2018learning}.

Addressing these challenges is crucial for several reasons. Effective anomaly detection systems can greatly enhance fraud detection, safeguarding insurance companies from significant financial losses and protecting policyholders from fraudulent activities. With an advanced anomaly detection system, insurance companies can identify fraudulent policyholders early, mitigating financial risk. Additionally, employing such systems fosters fairer and more sustainable business practices. Combining robust policy auditing tools with advanced anomaly detection systems can improve compliance with regulatory requirements and maintain higher operational standards. By focusing on unsupervised learning techniques, we aim to contribute valuable insights and tools for accurately detecting anomalies in the insurance industry.

In general, an unsupervised anomaly detection system works by identifying data points that deviate significantly from the norm, based solely on the intrinsic properties of the dataset. The challenge lies in defining the normal zones and clearly separating them from the anomalous points. Such systems make it easier to detect hidden patterns within the data, identify rare and unusual events, and uncover insights that might remain undetected using traditional methods.

This research aims to determine whether basic machine learning methods can effectively handle the complex task of anomaly detection within insurance contracts. A comprehensive survey of various anomaly detection techniques was conducted by \cite{chandola2009survey}, comparing several approaches. More recently, \cite{bauder2017medicare} evaluated different machine learning methods for detecting fraud or anomalies in medical care, including supervised, unsupervised, and hybrid methods. Their study explored techniques such as $k$-nearest neighbors (KNN), Mahalanobis distance-based approaches, and various autoencoders.

In our work, we conduct a detailed evaluation of basic machine learning and modern techniques, assessing their performance on datasets augmented with manually inserted anomalies. This allows us to benchmark their detection capabilities and evaluate their effectiveness with minimal human intervention. The methods examined include proximity-based, and tree-based approaches, which are traditionally favored for anomaly detection due to their simplicity and interpretability. Additionally, we explore deep learning techniques, such as autoencoders and variational autoencoders, to assess whether they offer more robust solutions for detecting anomalies, particularly in handling complex and high-dimensional data with minimal effort.

 It has been found that autoencoder-based reconstruction methods have been particularly effective in capturing non-linear correlations between features, thus handling non-linear relationships better than any traditional method (see \citet{zimmerer2018contextencoding}).  \cite{baur2019deep} demonstrated the efficiency of variational autoencoders for anomaly detection using reconstruction error. \cite{chen2018deep} explored the use of variational autoencoders and other deep generative models for anomaly detection in medical imaging, including brain tumor detection. While they noted the potential of these models, they highlighted limitations such as blurry reconstructions, which can hinder accurate lesion detection in datasets like BraTS 2015 (see \citet{BRATS2015}).

This research is divided into several parts. First, we will present two different datasets, which will be prepared for anomaly detection. In each dataset, we will introduce four anomalies to serve as indicators of how well the methods perform.

After preparing the data, we will apply two different types of methods -- classical machine learning methods and deep learning approaches -- to compare their performance and their ability to detect the aforementioned anomalies. To evaluate the performance of a few classical methods, we will introduce a primary scoring method that will enable the development of nearly fully autotuned tools.

Finally, we will compare the outcomes of these methods in terms of runtime, scoring accuracy, their ability to detect the four manually inserted anomalies in each dataset, and their tendency to avoid labeling too many points as anomalies. A research conducted by \cite{gomes2021insurance} in which basic machine learning models like KNN, SVM, Random Forest, and naive Bayes are compared with autoencoders showed that both simple autoencoders and variational autoencoders achieved higher accuracy in detecting anomalies, without the need for labeled data. Autoencoders have become one of the leading approaches for unsupervised anomaly detection in sequential data (see e.g. \citet{zimmerer2018contextencoding, chauhan2015anomaly, malhotra2016lstmbased}). This provides valuable insights into how the previously described methods can be applied to improve the quality of life insurance data.

\section{Data}
\label{Data-Discr}

To assess the goodness-of-fit and the effectiveness of various methods in detecting anomalies, suitable data is essential. The task is to develop a tool that will identify anomalous contracts within a life insurance dataset. However, finding such a dataset proves to be a major challenge. After some time spent searching, we decided to use two different open-source datasets. These datasets contain health insurance contracts, which, while not identical to life insurance contracts, are similar enough to allow for comparable analysis.

Dataset~1 comprises 986 observations, while Dataset~2 includes 25,000 observations. Although real-world tasks often involve datasets that are much larger and contain many more variables, these two datasets provide a useful comparison to highlight differences in the performance and overall behavior of the methods. It is important to note that both datasets are inherently unlabeled with respect to anomalies.

Dataset~1 \citep*{kaggle-Dataset-1} is relatively small, consisting of 986 rows and 12 columns. A summary and description of the columns in Dataset~1 can be found in Table~\ref{table:data_1_appendix_1} in the appendix. This dataset primarily contains basic customer information, premium price, and a few indicators regarding the presence of certain diseases. As mentioned, its complexity is low, which means even simple methods should be able to perform well and identify obvious anomalies.

In contrast, Dataset~2 \citep*{kaggle-Dataset-2} is considerably larger, with 25,000 rows and 24 columns. A detailed summary and explanation of the columns in Dataset~2 are provided in Table~\ref{table:dataset2} in the appendix. As previously mentioned, this dataset is more complex and significantly larger than Dataset~1. It includes more detailed information about the customers, their medical conditions, and aspects of their general lifestyle. While it is still much smaller than a typical life insurance dataset, it is sufficient to observe how the methods handle increasing size and complexity.

These open-source datasets were randomly selected and were not originally prepared for anomaly detection purposes, meaning no obvious outliers were present. To rigorously test each model's performance, we manually inserted four artificial anomalies into each dataset. These anomalies were carefully constructed to reflect a range of complexity, from easily identifiable cases to more subtle instances that closely mimic real-world anomalies.

The key aspect of these manually introduced anomalies is that they are designed with consideration for the overall structural characteristics of the data, rather than simple deviations in individual variables. This means that the anomalies represent unique or atypical combinations of features, simulating complex relationships found in real-world data where an entire record or observation might be anomalous due to the interplay between variables. By incorporating these structured anomalies, we can more accurately assess the ability of each model to identify both the intentionally introduced anomalies and any additional inherent outliers within the datasets. This approach allows for a nuanced comparative analysis of the models' sensitivity and robustness in detecting complex, realistic anomaly cases.

\section{Statistical methods}
\label{methods}
To accomplish the aforementioned objectives, various methods are required for effective anomaly detection. Below, we briefly introduce several classical machine learning techniques, along with two modern approaches known as autoencoders and variational autoencoders. We will also present scoring methods that enable us to evaluate the performance of several classical and deep learning approaches, and facilitate automated fine-tuning.

\subsection{Proximity-based methods}

Proximity-based approaches use the idea of distance to find anomalies. These approaches presume that normal data points are close to their neighbors and densely packed, whereas anomalies are far away. In this section, we will investigate several methods including some cluster-based methods and evaluate their usefulness in identifying abnormalities. All the methods were implemented using the \texttt{Python} package \texttt{scikit-learn} \citep{sklearn}.

\subsubsection*{Nearest-neighbors}
\label{NN}

The Nearest-Neighbor (NN) method is derived from the $k$-Nearest Neighbors (KNN) algorithm, which detects anomalies by evaluating the distance between data points and their nearest neighbors. In KNN, points far from their neighbors are flagged as anomalies, while points in denser regions, closer to their neighbors, are considered normal. The Nearest-Neighbor method modifies KNN to create an unsupervised approach, removing the reliance on labeled data (see \citet{NN_paper3}).

The KNN algorithm typically requires labeled data, where each data point belongs to a specific category. The parameter $k$ represents the number of neighbors to compare with the given point. For each data point, the algorithm calculates the distance to other points in the dataset using a distance metric, such as the Euclidean distance. Based on these distances, the $k$ nearest neighbors are identified, and KNN classifies the point based on the majority category of its neighbors.

In contrast, the Nearest-Neighbor method is designed for unsupervised anomaly detection by assigning an anomaly score to each point based on the distance to its $k$ nearest neighbors. This idea of calculating an anomaly score from the average distances to the $k$ nearest neighbors is introduced by \citet{NN_paper1}. Specifically, the score is computed by averaging the distances from a point to its $k$ nearest neighbors. Points with higher scores are considered more likely to be anomalies.

Once the anomaly scores for all points are calculated, a threshold is set. Points with scores exceeding this threshold are flagged as anomalies. In this work, the threshold is calculated using the mean and standard deviation of the nearest neighbor distances across the dataset.

\subsubsection*{$k$-means}

The $k$-means algorithm is a popular clustering method that divides data into $k$ groups by minimizing within-cluster variation. It assigns each data point to one of $k$ non-overlapping clusters, aiming to reduce differences within each cluster.

After clustering the data, each point's distance to its cluster centroid is calculated. Anomalies are identified if this distance exceeds a predefined threshold. This threshold is typically based on the average distance to the centroid, adjusted by the standard deviation of the distances, similar to using a normal distribution. Hence, this method is especially useful for detecting anomalies, as points that deviate significantly from the cluster centroids can be identified as outliers (see \citet{K-means_paper1, K-means_paper2}).

\subsubsection*{Density-based spatial clustering of applications with noise}
DBSCAN (Density-Based Spatial Clustering of Applications with Noise, see e.g. \citet{DBSCAN_paper1}) is a density-based clustering algorithm that identifies clusters of arbitrary shapes while detecting noise and outliers. It is particularly effective for large spatial datasets.

The algorithm operates by classifying points as core points, border points, or noise based on the number of neighbors within a defined radius $\varepsilon$. A point is a core point if it has at least $\kappa$ neighbors within its $\varepsilon$-neighborhood, initiating the formation of a cluster. Clusters are formed by recursively connecting density-connected points, while points that don't meet the core-point criteria but are close to core points are classified as border points. Any points that do not belong to any cluster are considered noise.

Choosing the parameters $\varepsilon$ and $\kappa$ is critical. The value of $\varepsilon$ can be determined using a $k$-distance plot, where the ``elbow" of the curve often indicates the optimal value. The parameter $\kappa$ is typically set slightly larger than the data's dimensionality to ensure clusters are dense enough to exclude noise effectively.

\subsubsection*{Hierarchical density-based spatial clustering of applications with noise}
\label{HDBSCAN}

HDBSCAN (Hierarchical Density-Based Spatial Clustering of Applications with Noise, see e.g. \citet{HDBSCAN-Advance-Knowledge}) is an enhancement of the traditional DBSCAN algorithm, introducing a hierarchical approach to better identify clusters in datasets with varying densities. The core idea of HDBSCAN is that dense regions in a dataset represent clusters, while less dense areas act as boundaries between them. Unlike DBSCAN, which operates with a fixed density threshold, HDBSCAN dynamically adjusts this threshold, creating a hierarchy of clusters that offers a more detailed analysis of the data structure.

A key feature of HDBSCAN is the introduction of the core distance, which reflects the local density around a point. This core distance is used to define the mutual reachability distance between points, which smooths out local density variations and prevents premature clustering of points with significantly different surrounding densities.

The choice of parameters in HDBSCAN, especially the minimum cluster size and $\kappa$, influences the density estimation and the cluster hierarchy. The minimum cluster size ensures that only robust clusters are identified, while $\kappa$ affects local density estimation, shaping the clustering hierarchy.

The extension from DBSCAN to HDBSCAN occurs by replacing DBSCAN's fixed density threshold with a hierarchical approach. HDBSCAN constructs a Minimum Spanning Tree (MST) based on mutual reachability distances, which considers both local density (via core distances) and direct point-to-point distances. This MST represents all possible clusters at various density levels. By progressively removing the highest-distance edges, HDBSCAN generates a hierarchical tree structure that captures clusters at different resolutions. A condensed tree summarizes this hierarchy, discarding small or insignificant clusters based on the minimum cluster size. The most stable clusters, which persist across a wide range of density thresholds, are then selected as the final output.

\subsubsection*{One-class support vector machine}

A One-Class Support Vector Machine (OCSVM, see e.g.  \citet{OneClassSVM}) is a machine learning technique specifically designed for anomaly detection. Unlike traditional support vector machines (SVMs, \citet{svm}) that are used for binary or multi-class classification, OCSVM focuses on distinguishing normal data points from anomalies. Instead of classifying between two or more classes, OCSVM creates a boundary that encloses the majority of data points, treating them as ``normal", while points outside this boundary are considered anomalies.

The core idea of OCSVM is to learn a function that classifies most of the data as normal (assigned a value of +1), while points outside this boundary are flagged as anomalies (assigned a value of -1). This is done by mapping the data into a high-dimensional feature space using a kernel function, such as the commonly used radial basis function (RBF) kernel (e.g., the Gaussian kernel, \citet{radial_kernel}), and then separating the data from the origin with a maximum margin. The result is a decision boundary that captures the majority of the data, allowing new points to be evaluated as either inside (normal) or outside (anomalous) this boundary.

One major difference between OCSVM and traditional SVMs is that OCSVM is unsupervised, meaning it does not require labeled training data. Instead, it assumes that most of the data points are normal and focuses on identifying outliers. The balance between the boundary tightness and the number of points considered anomalies is controlled by a parameter, $\nu$, which determines the fraction of points allowed to lie outside the boundary.

\subsection{Tree-based anomaly detection methods}

Tree-based methods utilize decision trees to identify anomalies. These methods are effective in handling high-dimensional data and can provide interpretable results. In this section, we will discuss in detail how the Isolation Forest algorithm works.

\subsubsection*{Isolation Forest}

Isolation Forest (iForest) is an anomaly detection algorithm that identifies anomalies by isolating data points. Following \citet{IsolationForest}, the core idea is that anomalies are easier to isolate because they typically lie in low-density regions or are far from other points, making their separation simpler.

The algorithm works by constructing multiple Isolation Trees (iTrees). Each tree recursively partitions the data by selecting random features and random split points, isolating data points at the leaf nodes. Anomalies are usually isolated faster (i.e., at shallower depths in the tree) due to their distinct characteristics, while normal points require more partitions for isolation.

Anomaly detection in iForest is based on the average depth at which a point is isolated across all trees. A point with a shallow average isolation depth is considered an anomaly, while points requiring deeper partitions are classified as normal. The anomaly score ranges from $-1$ to $1$, with a score near $1$ indicating a high likelihood of a normal point, and a score near $-1$ indicating a high likelihood of an anomaly. Typically, points with negative scores are classified as anomalies.

For implementation, we used the \texttt{IsolationForest} function from the \texttt{scikit-learn} package (Version 1.5.0, \citet{sklearn}). This function employs a heuristic that identifies points likely to be anomalies by ranking them based on their anomaly scores. 

\subsection{Deep learning-based anomaly detection methods}

In this section, we will discuss two types of unsupervised deep learning methods that use the power of neural networks to detect anomalies. These methods can find complex patterns and relationships within any dataset, making them powerful tools in the field of anomaly detection. Furthermore, they can process large and complex datasets that traditional methods might struggle with. \texttt{PyTorch} \citep{pytorch}, an open-source deep learning framework, was used to build the models for both methods.

\subsubsection*{Autoencoder}

An autoencoder (AE) is a neural network used to learn efficient data representations by compressing high-dimensional data into a lower-dimensional latent space and then reconstructing the original data from it. \citet{Autoencoder} explains that the goal is to capture meaningful features of the dataset by minimizing the difference between the input and the reconstructed output, indicating that the model has learned the underlying patterns of the data.

The AE consists of three main components: an encoder, which compresses the input data into a reduced latent space; a latent feature representation, which acts as a bottleneck forcing the network to retain the most important data features; and a decoder, which reconstructs the data from this compressed form. The encoder compresses the data from its original dimension $d$ to a lower dimension $p$, while the decoder reconstructs the input to its original dimension.

In this work, a Feed-Forward Autoencoder (FFA, see also \citet{Autoencoder}) is used, which processes data in a single direction, from input to output, without recurrent connections or feedback loops. This architecture contrasts with more complex variations like Recurrent Autoencoders (for time-series data, \citep{Rec_AE}) and Convolutional Autoencoders (for image data, \citep{conv_AE}). FFAs are typically used for simpler, tabular data. Both the encoder and decoder may consist of multiple layers, with the encoder gradually reducing the dimensionality until it reaches the bottleneck, and the decoder symmetrically expanding the data back to its original dimension.

To optimize the model, the Mean Squared Error (MSE) loss function is used, measuring how well the AE reconstructs the input. A low reconstruction error indicates that the AE has captured the essential data patterns, while a high error suggests difficulties in learning these patterns or the presence of anomalies. For anomaly detection, a threshold is set on the reconstruction error: if the error exceeds this threshold, it signals potential anomalies. In this work, normalization was applied to the input data, as it helps improve convergence and enhances the autoencoder's ability to minimize the MSE and effectively learn the dataset's structure.

\subsubsection*{Variational autoencoder}

In the previous section, we described how autoencoders (AEs) work. Variational autoencoders (VAEs) are an extension of classical AEs. While the core idea is similar, the key difference is that, in VAEs, the latent space is not just a lower-dimensional representation. Instead, \citet{VAE-paper} shows that VAEs model the latent variables as being drawn from a normal distribution.

This modification introduces probabilistic encoding and decoding, in contrast to the deterministic nature of classical AEs. The encoder reduces the input data to a smaller latent space, and instead of representing the latent variables directly, the VAE estimates the mean and variance of a normal distribution from which the latent variables are sampled. The decoder then uses these sampled values to reconstruct the original data. Similar to AEs, VAEs aim to minimize the difference between the original data and its reconstruction, while also ensuring the latent space is well-structured.

The VAE's objective function includes two parts: the reconstruction loss, which measures how well the VAE can reconstruct the input data, and the Kullback-Leibler (KL) divergence (see e.g. \citet{KL}), which regularizes the latent space by ensuring the learned distribution is close to a standard normal distribution. This helps maintain a smooth and interpretable latent space, where meaningful data generation is possible.

To optimize the model, the reparametrization trick is applied, making the sampling process differentiable and enabling gradient-based optimization. Handling anomalies in VAEs is similar to AEs, where a threshold is used to detect outliers based on reconstruction error. Normalization of the input data is also applied in VAEs to improve model performance, ensuring better convergence and a balanced trade-off between accurate reconstruction and a well-regularized latent space. 

\subsection{Scoring methods}

Finally, there are various scoring methods that allow data scientists to assess and compare the goodness-of-fit of models.

\subsubsection*{Silhouette Score}
The Silhouette Score is a metric used to evaluate the quality of clustering by measuring how well each data point fits within its own cluster compared to other clusters. For each point, it calculates two main values: the average distance to points within the same cluster and the average distance to the nearest neighboring cluster. The Silhouette Score ranges from -1 to 1, where a score close to 1 indicates good clustering, and a score near -1 suggests poor clustering.

The overall Silhouette Score for a dataset is obtained by averaging the scores of all individual points, providing a measure of clustering quality. This metric not only evaluates clustering performance but also penalizes attempts to include outliers in clusters, making it useful for anomaly detection (see \citet{Silhouette_score}).

\subsubsection*{Anomaly Score}

In an Isolation Forest, the Anomaly Score measures how ``anomalous" a data point is by evaluating how easily it can be isolated from the rest of the dataset (see \citet{IsolationForest}). The method works by recursively splitting the data using randomly selected features and split values until each point is isolated. Anomalous points, which tend to be more isolated and located in sparser regions, require fewer splits to separate them from the dataset. The anomaly score reflects this, with more negative scores indicating a higher likelihood of being an anomaly, while higher positive scores suggest that a point is more likely to be normal. This approach helps to effectively identify and rank data points for unsupervised anomaly detection.

\subsubsection*{Reconstruction Error}

Reconstruction error is a key concept in autoencoders (see \citet{Autoencoder}), representing the difference between the original input data and the reconstructed output produced by the decoder from the compressed representation (latent space). The goal of an autoencoder is to recreate the input as accurately as possible, hence the reconstruction error quantifies how well or poorly this is achieved. A low reconstruction error indicates that the reconstruction closely matches the original input, while a high error suggests that the autoencoder could not reproduce the input well. In applications such as anomaly detection, a high reconstruction error can signal that the data is anomalous, as the autoencoder finds it difficult to reconstruct.

\section{Statistical analysis}
After the introduction of the datasets and the methods, we will discuss the results of this anomaly detecion application. Therefore, we will also discuss preprocessing techniques used on these datasets. In the end, we will compare the results of all the models and discuss which method can detect all four manually placed anomalies and which is best suited for anomaly detection in the life and health insurance domains, respectively.

All models discussed in this section are trained locally on a machine equipped with an Intel Core i5 13th generation processor and 16 GB of RAM. The software tools utilized for this analysis include \texttt{Python} \citep{python}, the programming language used to code all the methods, \texttt{Pandas} \citep{pandas}, a data analysis library to create data frames for easy data manipulation, \texttt{NumPy} \citep{numpy}, a fundamental package for numerical computation, matrix multiplications, etc., \texttt{tqdm} \citep{tqdm}, a fast and extensible progress bar, \texttt{kneed} \citep{kneed}, a library for locating the knee point on a curve and \texttt{Matplotlib} \citep{matplotlib}, a library for creating both static and interactive visualizations.

\subsection{Data cleaning and preprocessing}

Data cleaning and preprocessing are critical steps required for training machine learning and deep learning models, and for ensuring that these models are free from bias. In this subsection, we will discuss all the data preprocessing methods used in the work for model training.

The smaller dataset, Dataset~1, has $12$ continuous columns and 986 observations with no missing values. Since the dataset comprises variables $ Height$ and $ Weight$, we decided to add a new column $BMI$, which is an essential parameter in the health insurance industry and is calculated by $\textit{BMI} = \frac{\textit{Weight}}{\textit{(Height)}^2}$. This increased the count of columns from $12$ to $13$. Additionally, four anomalous observations are added to the dataset, increasing the observation counts from $986$ to $990$. It is important to note that these anomalous observations may contain unusual values in all or a few fields.

The more extensive dataset, i.e., Dataset~2, has $24$ columns, of which $13$ are categorical, $11$ are continuous, and $25,000$ observations. The dataset contains numerous observations with missing $BMI$ values, and there are also a significant number of missing entries in the column ``Year$\_$last$\_$admitted" that must be addressed prior to model training. This, in total, removed $994$ observations, reducing the dataset size from $25,000$ to $24,006$. The descision to remove those observations with missing values was made to ensure the integrity and accuracy of the analysis. The reduction of approximatly $4\%$ of the dataset still left a dataset which was big and complex enough to compare the results with Dataset~1. These missing values can introduce bias, reduce the statistical power of the dataset, and lead to incorrect conclusions. By excluding these observations, we ensure the dataset is complete and the results are reliable. Since the datasets were not originally prepared for anomaly detection, removing observations with missing values does not result in substantial informational loss. Ensuring data completeness by eliminating entries with missing values helps maintain consistency and reliability, without compromising the integrity of the overall analysis. This allows for a more accurate comparison of model performance and reduces potential noise that could impact the results. Additionally, all categorical variables in the datasets are transformed using one-hot encoding. This method converts categorical variables into a set of binary variables, where each new variable represents one category from the original data. A value of $1$ indicates the presence of the corresponding category, while $0$ indicates its absence (see \citet{one-hot}). As a result, the dimensionality of the dataset increases from $24$ to $51$ columns. Additionally, four anomalous observations are added to the dataset, increasing the observation counts from $24,006$ to $24,010$.

A preliminary analysis is carried out by creating models with and without scaling the datasets, and the decision is made to train classical models on unscaled datasets to get a feeling of the original data. On the other side, scaled datasets are used to train autoencoders and variational autoencoders which only can handle such type of data. Here, \texttt{StandardScaler()} function from the \texttt{scikit-learn} library is used to scale both the datasets.
\subsection{Classical anomaly detection methods}
We applied several classical anomaly detection methods presented in Section~\ref{methods} to the two datasets from Section \ref{Data-Discr}, and their results are shown in Table \ref{tab:basic}.  Further details on model parameters as well as detailed results are discussed in Section~\ref{ModParams} of the appendix.

\begin{table}[h!]	
	\centering
	\resizebox{\textwidth}{!}{
	\begin{tabular}{|l|c|c|c|c|c|c|}
	\hline
	 & \multicolumn{3}{c|}{\textbf{Dataset~1 (990 Rows)}} & \multicolumn{3}{c|}{\textbf{Dataset~2 (24,010 Rows)}}                                               \\ \hline
		& \textbf{\begin{tabular}[c]{@{}c@{}}Anomaly\\ detected\\ out of 4\end{tabular}} & \makecell{\textbf{Total} \\ \textbf{runnung} \\ \textbf{time}} & \makecell{\textbf{Silhouette} \\ \textbf{Score}} & \textbf{\begin{tabular}[c]{@{}c@{}}Anomaly\\ detected\\ out of 4\end{tabular}} & \makecell{\textbf{Total} \\ \textbf{runnung} \\ \textbf{time}} & \makecell{\textbf{Silhouette} \\ \textbf{Score}} \\
	\hline 
	NN         & $3$                              &  $<1$sec                & -                & $2$                              & $1.1$sec                   & -                \\ \hline 	
	$k$-means                      & $3$                              & $1:05$min                        & $0.95$             & $0$                              & $2.5$h                 & $0.57$             \\ \hline 	 
	DBSCAN                     & $\textbf{4}$                              & $6.7$sec                         & $0.93$             & DNF                           & $>5$h              & -                \\ \hline 	
	HDBSCAN     & $\textbf{4}$                              & $4.6$sec                         & $0.91$             & DNF                           & $>5$h                 & -                \\ \hline 	
	OCSVM  & $2$                              & $<1$sec                & -                & $1$                              & $13.5$sec                  & -                \\	\hline 	
	Isolation Forest           & $\textbf{4}$                              & $<1$sec                 & $0.48$**           & $\textbf{4}$                              & $8.5$sec                   & $0.46$**           \\ 
 \hline
		\end{tabular}
	}
	\caption{Comparison of various classical anomaly detection methods.}
		\label{tab:basic}
	\begin{tablenotes}
		\small
		\centering
		\item[1] (\textit{DNF}: \textit{Did Not Finish}.)
		\item[2] (\textit{**} \textit{Average Anomaly Score}.) 
	\end{tablenotes}
	\end{table}

Table~\ref{tab:basic} summarizes the runtime (RT), the number of detected anomalies from the four manually inserted anomalies, and the respective scoring metrics. To ensure usability for non-data scientists, methods that allow fully automatic parameter tuning, such as Isolation Forest, $k$-means, and DBSCAN, were trained using the Silhouette Score (or Anomaly Score in the case of Isolation Forest) to achieve optimal performance. A grid search was employed to facilitate this tuning process, utilizing the Silhouette Score or Anomaly Score to identify and select the best model seamlessly. For these methods, a grid search was employed for automatic parameter tuning. The remaining methods were manually adjusted to detect as many of the four manually inserted anomalies as possible. In the case of HDBSCAN, although it provides a Silhouette Score, parameter selection was guided by exploratory data analysis of the dataset. As mentioned in \citet{HDBSCAN_autotune} and explained in Section \ref{HDBSCAN}, the advantage of HDBSCAN is the flexibility it provides in capturing a hierarchy of clusters. Since different datasets have widely varying structures, the ``optimal" parameters for one dataset may not generalize well to others. This, combined with the exploratory nature of parameter selection, results in a vast number of potential parameter settings. Consequently, it becomes impractical to autotune the parameters of this method efficiently.

The results indicate that more advanced methods are required, as these classical approaches demonstrate significant limitations, especially when applied to larger and more complex datasets like Dataset~2. Most methods handled the smaller Dataset~1 rather well, detecting between two and four of the manually inserted anomalies. However, methods such as NN, $k$-means, and OCSVM struggled to detect all anomalies. While Isolation Forest successfully detected all four anomalies, it overall labeled $342$ of $990$ points as anomalies, resulting in a low Anomaly Score of $0.48$.

As the dataset size and complexity increased, these limitations became more apparent. DBSCAN and HDBSCAN failed to complete within five hours. $k$-means took around $2.5$ hours and detected only two out of four anomalies, reflected by its low Silhouette Score of $0.57$. NN and OCSVM, despite their faster runtimes, performed even worse, detecting only two and one anomalies, respectively. Although Isolation Forest managed to detect all four anomalies within a reasonable runtime of $8.5$ seconds, its low Anomaly Score of $0.46$ casts doubt on the model's reliability.

\subsection{Deep learning-based anomaly detection methods}
Following the presentation of the results from various classical methods for anomaly detection applied to the two datasets, we will discuss the outcomes of the deep learning-based anomaly detection techniques. In this section, we will focus on (AEs) and (VAEs).

The main challange building AE and VAE models is the choice of the dimensions of all the layers. Because we have no labels in the datasets we have to use the methods in an unsupervised way. The probleme here is that there is no guarantee that the final model and its output are good. It is possible that we may select a model architecture that results in the poorest performance. Therefore, we use the idea of ``ensamble learning". To proceed this, we choose a number $M$ of models which has to be trained. All these models have different architectures. The only measure we have is the reconstruction error. After getting the reconstruction error of each model for each point of the given dataset, we can build an average of these reconstruction errors. Finally, there is an average reconstruction error for each point. From now on we can choose a threshold $t$ above which all the reconstruction errors indicate the presence of anomalies.

In this work, we built three different models for each method, applying them to the two given datasets. The architectures were chosen experimentally to explore a variety of configurations, including both extreme and intermediate values for the network's hidden layers and latent space. Specifically, while the input layer remained consistent across all models, the sizes of the hidden layers and latent space varied: one configuration had very large layers, another had medium-sized layers, and the third had very small ones. This approach was designed to assess the impact of different architectural choices and reduce the risk of selecting suboptimal models due to random architecture selection. The final settings as well as the detailed results can be seen in Section~\ref{AE-settings} in the appendix. Finally, the results of the models are presented in Table \ref{tab:DL-method}.

\begin{table}[h!]
	\centering
	\resizebox{\textwidth}{!}{

	\begin{tabular}{|l|c|c|c|c|c|c|}
	\hline
	                 & \multicolumn{3}{c|}{\textbf{Dataset~1 (990 Rows)}}                                                                 & \multicolumn{3}{c|}{\textbf{Dataset~2 (24,010 Rows)}}                                                                \\ \hline
									   & \textbf{\begin{tabular}[c]{@{}c@{}}Anomaly\\ detected\\ out of 4\end{tabular}} & \makecell{\textbf{Total} \\ \textbf{runnung} \\ \textbf{time}} & \makecell{\textbf{Total} \\ \textbf{models}} & \textbf{\begin{tabular}[c]{@{}c@{}}Anomaly\\ detected\\ out of 4\end{tabular}} & \makecell{\textbf{Total} \\ \textbf{runnung} \\ \textbf{time}} & \makecell{\textbf{Total} \\ \textbf{models}} \\ \hline
	Autoencoder                        & $4$                              & $2:22$min                        & $3$                      & $4$                              & $6$min                           & $3$                      \\  \hline 	
	Variational autoencoder            & $4$                              & $4:35$min                        & $3$                      & $4$                              & $20$min                          & $3$                      \\ \hline
	\end{tabular}
	}
	\caption{Comparison of autoencoder and variational autoencoder ensemble.}
	\label{tab:DL-method}
	\end{table}

First of all, we have to mention that all the models needed an acceptabe time for been trained. The methods were not particularly fast on the small dataset, taking approximately two to five minutes to complete. On the other hand, the runtime doesn't increase much on the second dataset, which is a big difference to the classical methods. We obtained similar findings with respect to the performance. The AEs as well as the VAEs were able to detect all the four manually placed anomalies in the both datasets. The only notable problem is that the VAE marked over $50\%$ of the observations from Dataset~1 as anomalies which isn't a trustable result (see Table \ref{tab:AutoEncoder_table}). For AEs on both datasets and for VAE on the second dataset the normal points to anomalies ratio was very moderat.

Based on these results, it is evident that the ensemble of autoencoders surpasses all the basic machine learning methods discussed in this work, making it a strong choice for anomaly detection tasks. The methods not only performed better in terms of accuracy, but the runtime on both datasets is optimal.

\subsection{Comparison of the methods}

To summarize the results, we will examine the ability of the methods to detect ``real" anomalies while minimizing the number of observations falsely marked as anomalies.
\begin{table}[h!]
	\centering
	\resizebox{\textwidth}{!}{

	\begin{tabular}{|l|c|c|}
	\hline
	\textbf{Methods}                   & \textbf{Dataset 1 (990 Rows)}                                                                 & \textbf{Dataset 2 (24,010 Rows)}                                                                \\ \hline
NN & $14$ & $395$ \\ \hline 
$k$-means & $50$ & $1201$ \\ \hline
DBSCAN & $38$ & DNF \\ \hline 
HDBSCAN & $22$ & DNF \\ \hline
OCSVM & $83$ & $2401$ \\ \hline
Isolation Forest & $\textbf{342}$ & $1974$ \\ \hline
	Autoencoder                        &    $224$ & $705$                  \\ \hline
	Variational autoencoder            & $\textbf{512}$ & $548$ \\ \hline
	\end{tabular}
	}
	\caption{Number of total detected anomalies by each method.}
	\label{tab:total_anomalies}
	\begin{tablenotes}
		\small
		\centering
		\item[1] (\textit{DNF}: \textit{Did Not Finish}.)
	\end{tablenotes}
	\end{table}
Table~\ref{tab:total_anomalies} shows that classical methods are generally able to mark only a small number of observations as anomalies in Dataset~1. AEs, VAEs, and Isolation Forest, on the other hand, tend to mark between $20\%$ and $55\%$ of the dataset as anomalous. This is a concerning result, as it suggests these methods may not be effectively understanding the underlying structure of the data. Instead, they appear to ``guess", marking many points as anomalies to avoid potential mistakes, likely because they only capture the structure of a few data points.

In contrast, for the larger Dataset~2 the picture changes significantly. While some classical methods were unable to finish processing within an acceptable timeframe, it is evident that the deep learning methods successfully identified clear patterns in the data, marking only a small percentage of observations as anomalies. On the other hand, the classical methods that did finish, with the exception of NN, marked significantly more points as anomalies compared to the deep learning approaches.

\begin{figure}[htp]
\centering
\includegraphics[scale=0.65]{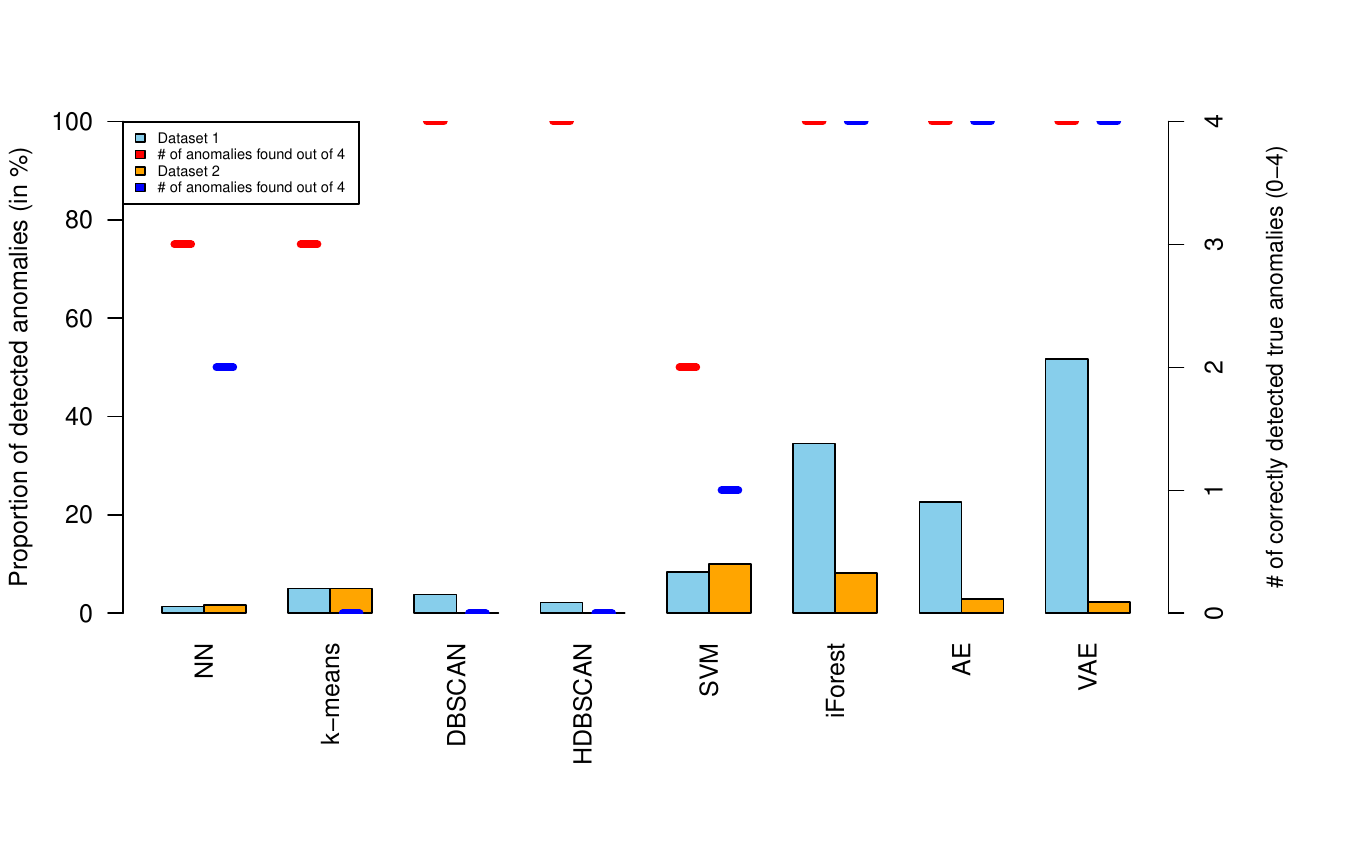}
\caption{Proportion of anomalies and detected manual anomalies (secondary $y$-axis).}
\label{fig:anomalies_percent_manual}
\end{figure}

When combined with the ability of these methods to detect the four manually inserted anomalies, Figure~\ref{fig:anomalies_percent_manual} provides a clear picture of the findings from the analysis. The bars represent the percentage of observations marked as anomalous by each method, as shown in Table~\ref{tab:total_anomalies}. The blue and red lines indicate the number of manually inserted anomalies detected by each method for each dataset. In Dataset~1, we observe similar performance among DBSCAN, HDBSCAN, Isolation Forest, AE, and VAE in detecting the four manually inserted anomalies. The key difference is that the deep learning methods mark many more observations as anomalies compared to the classical methods (except for Isolation Forest). Dataset~2 supports the observation made in previous sections, showing that only the deep learning methods and Isolation Forest were able to detect all four manually placed anomalies. However, the issue with Isolation Forest is that it marks more than twice as many observations as the deep learning methods, highlighting that the modern ensemble methods tend to provide more reliable results than the classical ones.

\section{Conclusion}

This study explores various machine learning and deep learning-based anomaly detection methods to help insurance companies identify irregularities in their data. The goal of the research was to develop an unsupervised anomaly detection method that could accurately detect anomalies in datasets of any size, while automatically tuning hyperparameters based on the input data. The primary focus was on life insurance, which posed challenges due to the lack of available datasets.

To conduct the planned analyses, we opted for alternative, open-source datasets closely related to life insurance, selecting two health insurance datasets instead. The first dataset is smaller, with $986$ rows (contracts) and $12$ columns (variables), containing limited customer information. In contrast, the second dataset is much larger, with $25,000$ rows and $24$ columns, representing more complex contracts. Dataset~2 also includes categorical variables, which we transformed using one hot encoding, significantly increasing the number of columns. Additionally, Dataset~2 contained some contracts with missing data, which we decided to remove to ensure smoother ana\-lyses. To validate the performance of the anomaly detection methods, we manually introduced four anomalous contracts into each dataset.

After preprocessing the datasets, we trained several models and compared their outputs. Among the classical proximity-based anomaly detection methods were Nearest Neighbors, $k$-means, DBSCAN, HDBSCAN, and One-Class SVM. We also trained an Isolation Forest, representing tree-based methods. These were compared against autoencoders (AEs) and variational autoencoders (VAEs). Since AEs and VAEs lack a direct scoring mechanism, we used ensemble learning with three models in each case. Where applicable, the classical methods were evaluated using the Silhouette Score, while the Isolation Forest was rated using the Anomaly Score, enabling us to automate hyperparameter tuning in some instances.

The comparison of classical methods yielded mixed results. Most methods performed reasonably well on Dataset~1, detecting between two and four of the manually inserted anomalies. However, on Dataset~2, which is larger and more complex, nearly every method failed. DBSCAN and HDBSCAN were unable to complete, and apart from the Isolation Forest, the other methods produced poor results. The Isolation Forest detected all four anomalies but was hindered by its Anomaly Score, which fell below $0.5$, indicating a lack of trustworthiness. In contrast, the ensembles of AEs and VAEs successfully detected all the anomalies in both datasets. Their runtime was acceptable, with Dataset~2's size increase remaining within a manageable range. However, VAEs struggled on the smaller dataset, flagging over $50\%$ of the data as anomalous.

In conclusion, the results suggest that an ensemble of autoencoders provides higher accuracy for anomaly detection. However, this approach is computationally expensive and requires a powerful processor or GPU for efficient processing -- though this should not be an issue for insurers, who can afford to run such models over weekends. Additionally, the Isolation Forest emerged as the best-performing classical method, offering a balance between accuracy and efficiency. While it detected all four anomalies in both datasets within an optimal time frame, its low average Anomaly Score raised concerns about its reliability.

\subsection*{Recommendations for future research}

Future research in the field of anomaly detection techniques should include a number of ways to improve the resilience and scalability of current systems. One approach aims to optimize and fine-tune traditional methods to enhance their robustness and scalability, potentially providing a more effective and efficient approach to anomaly detection across diverse datasets. Additionally, exploring hybrid models that integrate the strengths of both traditional and deep learning methods could yield improved performance and efficiency in anomaly detection.

Expanding the use of ensembles, especially with autoencoders and variational autoencoders, offers another promising direction. Instead of limiting ensembles to three models, researchers could experiment with ensembles of five, ten, or more models to evaluate if larger ensembles lead to better anomaly detection results. Moreover, combining different types of models within an ensemble, such as five autoencoders and five variational autoencoders, may leverage the strengths of both approaches, potentially enhancing accuracy and robustness. It would also be interesting to experiment with different ways of combining model outputs. On the one hand, the outputs of each model can be analyzed separately and combined afterwards. On the other hand, models can be directly combined and their collective output analyzed, as done in this work. Furthermore, exploring alternative methods for combining outputs could further improve performance.

Moreover, investigating hybrid models that incorporate basic machine learning methods like Isolation Forests, neural networks, or $k$-means as a preprocessing step before applying autoencoders or variational autoencoders could help filter out obvious abnormalities. This would reduce computational costs while maintaining strong anomaly detection accuracy. Moreover, by optimizing and fine-tuning traditional approaches, they could become more scalable and better suited for larger datasets.

In addition, researchers might explore running the models on powerful systems or GPUs, which could greatly reduce runtime and enable faster experimentation and analysis, particularly when working with large datasets or complex models. This optimization can improve the performance of anomaly detection systems, making them more suitable for real-world applications.

\pagebreak
\begin{appendix}
\section{Appendix}\label{sec:append}
\subsection{Data}
\begin{table}[htp]
	\centering
	\begin{tabular}{|l|l|p{6cm}|} 
	\hline
		\textbf{Variable name} & \textbf{Variable type} & \textbf{Description} \\ \hline
		\textit{ID} & Categorical & Unique identifier for each entry \\ \hline
		\textit{Age} & Continuous & Age of the individual \\ \hline
		\textit{Diabetes} & Categorical & Indicates if the individual has diabetes \\ \hline
		\textit{BloodPressureProblems} & Categorical & Indicates if the individual has blood pressure problems \\ \hline
		\textit{AnyTransplants} & Categorical & Indicates if the individual has had any transplants \\ \hline
		\textit{AnyChronicDiseases} & Categorical & Indicates if the individual has any chronic diseases \\ \hline
		\textit{Height} & Continuous & Height of the individual \\ \hline
		\textit{Weight} & Continuous & Weight of the individual \\ \hline
		\textit{KnownAllergies} & Categorical & Indicates if the individual has known allergies \\ \hline
		\textit{HistoryOfCancerInFamily} & Categorical & Indicates if there is a history of cancer in the family \\ \hline
		\textit{NumberOfMajorSurgeries} & Continuous & Number of major surgeries the individual has had \\ \hline
		\textit{PremiumPrice} & Continuous & Price of the insurance premium \\ \hline
	\end{tabular}
	\caption{Description of variables in Dataset~1 \citep*{kaggle-Dataset-1}.}
	\label{table:data_1_appendix_1}
\end{table}

\begin{table}[htp]
	\centering
	\begin{tabular}{|l|p{2cm}|p{6cm}|} 
	\hline
		\textbf{Variable Name} & \textbf{Variable Type} & \textbf{Description} \\ \hline
		\textit{applicant\_id} & Categorical & Unique identifier for each individual \\ \hline
		\textit{years\_of\_insurance\_with\_us} & Continuous & Number of years the individual has been insured \\ \hline
		\textit{regular\_checkup\_last\_year} & Continuous & Whether the individual had a regular checkup last year \\ \hline
		\textit{adventure\_sports} & Categorical & Indicates if the individual participates in adventure sports \\ \hline
		\textit{Occupation} & Categorical & Individual's occupation \\ \hline
		\textit{visited\_doctor\_last\_1\_year} & Continuous & Number of doctor visits in the last year \\ \hline
		\textit{cholesterol\_level} & Categorical & Individual's cholesterol level \\ \hline
		\textit{daily\_avg\_steps} & Continuous & Average number of steps taken daily \\ \hline
		\textit{age} & Continuous & Individual's age \\ \hline
		\textit{heart\_disease\_history} & Categorical & Indicates if there is a history of heart disease \\ \hline
		\textit{other\_major\_disease\_history} & Categorical & Indicates if there is a history of other major diseases \\ \hline
		\textit{Gender} & Categorical & Individual's gender \\ \hline
		\textit{avg\_glucose\_level} & Continuous & Average glucose level \\ \hline
		\textit{bmi} & Continuous & Body Mass Index (BMI) \\ \hline
		\textit{smoking\_status} & Categorical & Smoking status \\ \hline
		\textit{Year\_last\_admitted} & Categorical & Year when the individual was last admitted to a hospital \\ \hline
		\textit{Location} & Categorical & Individual's location \\ \hline
		\textit{weight} & Continuous & Individual's weight \\ \hline
		\textit{covered\_by\_any\_other\_company} & Categorical & Indicates if the individual is covered by any other insurance company \\ \hline
		\textit{Alcohol} & Categorical & Alcohol consumption status \\ \hline
		\textit{exercise} & Categorical & Exercise habits \\ \hline
		\textit{weight\_change\_in\_last\_one\_year} & Continuous & Change in weight over the last year \\ \hline
		\textit{fat\_percentage} & Continuous & Body fat percentage \\ \hline
		\textit{insurance\_cost} & Continuous & Cost of insurance \\ \hline
	\end{tabular}
	\caption{Description of variables in Dataset~2 \citep*{kaggle-Dataset-2}.}
	\label{table:dataset2}
	\end{table}
\pagebreak

\subsection{Model Settings and Results}
\subsubsection{Classical Methods}\label{ModParams}
\subsubsection*{Nearest-Neighbors}

\begin{itemize} 
\item No scoring mechanism $\Rightarrow$ tuning $k$ by selecting the model that detects the most of the four manually placed anomalies within the dataset 
\item A range for $k$ is defined, i.e., $k \in \lbrace 2, \ldots, 100 \rbrace$ 
\item For Dataset~1, $k=3$ is empirically the best choice  
\item Due to the lack of automatic parameter tuning, $k=3$ was also chosen for Dataset~2  
\end{itemize}

\subsubsection*{$k$-means}

\begin{itemize} 
\item Method can be automatically fine-tuned 
\item For Dataset~1, the best model has $k=19$
\item Dataset~2 delivers poor results: the best model is for $k=2$
\item Additionally, very long runtime for Dataset~2, approximately $2.5$ hours 
\end{itemize}

\subsubsection*{DBSCAN}

\begin{itemize} 
\item Automatic parameter tuning is possible, conducted via grid search using the function \texttt{kneeLocator()}, optimizing the Silhouette Score of the model 
\item The main parameters are $\varepsilon$ and $\kappa$ 
\item Performance on Dataset~1 after tuning is very good
\item Unable to complete on Dataset~2
\end{itemize}

\subsubsection*{HDBSCAN}

\begin{itemize} 
\item Lack of automation techniques \citep{HDBSCAN_autotune} $\Rightarrow$ parameters need to be determined experimentally 
\item Main parameters are $\kappa$ and ``min cluster size", where $\kappa$ can be borrowed from the best DBSCAN models; min cluster size is found by manually comparing the results 
\item Better performance on Dataset~1 than DBSCAN 
\item Similarly to DBSCAN, it was unable to finish on Dataset~2
\end{itemize}

\subsubsection*{OCSVM}

\begin{itemize} 
\item Cannot be automatically tuned 
\item Similar approach to NN $\rightarrow$ best model for Dataset~1 is found experimentally and applied to Dataset~2 
\item Main parameter is $\nu$; the kernel used is the radial basis function (RBF) as commonly suggested in the literature \citep{radial_kernel}
\item The best model on Dataset~1 has $\nu=0.1$; the same model performs worse on Dataset~2 
\end{itemize}

\subsubsection*{Isolation Forest}

\begin{itemize} 
\item Anomaly Score $ \rightarrow$ automatic parameter tuning using grid search  
\item Main parameters: maximum number of features considered for splitting $f\_max$, maximum number of samples drawn to build each tree $s\_max$, and the number of base estimators in the ensemble $n\_est$ 
\item For both datasets, the best models were identical, and the results were similar (see Table \ref{tab:Res-table-dat1_2})
\end{itemize}

\subsubsection*{Summary}
\begin{itemize}
\item Table \ref{tab:Res-table-dat1_2} summarizes all the settings and results of the classical methods applied on both datasets
\item All results are produced as described in the preceding paragraphs
\end{itemize}

\begin{table}[!h]
	\centering
	\resizebox{\textwidth}{!}{
	\begin{tabular}{|l|c|c|c|c|c|c|}
	\hline
\textbf{Dataset~1} &    &    &    &  & & \\ \hline
	& NN & $k$-means &DBSCAN &HDBSCAN & OCSVM & iForest \\
	\hline 
 
$k$ & $3$   & $19$   &    &  & & \\ \hline
$t$ & $292.52$   &     &     &  &  & \\ \hline
Total anomalies & $14$   & $50$   & $38$    & $22$ & $83$ & $342$\\ \hline
Anomaly out of $4$ & $3$   & $3$   & $4$    & $4$ & $2$ &$4$ \\ \hline
RT & $<1$sec   & $1:05$min   & $6.7$sec    & $4.6$sec & $<1$sec & $<1$sec \\ \hline
Silhouette Score &  &$0.95$  & $0.91$    & $0.93$ &  & \\ \hline
$\varepsilon$ &    &    & $1000$    &  &  & \\ \hline
$\kappa$ &    &    & $10$    & $10$ &  & \\ \hline
Min cluster size &    &    &     & $5$ &  & \\ \hline
$\nu$ &    &    &     &  & $0.1$ & \\ \hline
kernel &    &    &     &  & RBF & \\ \hline
$f\_max$ &    &    &     &  &  &$1.0$ \\ \hline
$s\_max$ &    &    &     &  &  &$0.5$ \\ \hline
$n\_est$ &    &    &     &  &  &$50$ \\ \hline
Anomaly Score &    &    &     &  &  &$0.48$ \\ \hline 
            \hline 
\textbf{Dataset~2}          & & & & &  &  \\
           \hline 
$k$ & $3$   & $2$   &    &  & & \\ \hline
$t$ & $101.12$   &     &     &  &  & \\ \hline
Total anomalies & $395$   & $1201$   & DNF    & DNF & $2401$ & $1974$\\ \hline 
Anomaly out of $4$ & $2$ &$0$  & DNF   & DNF    & $1$ &$4$ \\ \hline
RT & $1.1$sec   & $2.5$h   & $>5$h    & $>5$h & $8.5$sec & $8.5$sec \\ \hline 
Silhouette Score &  &$0.57$  & DNF    & DNF &  & \\ \hline
$\varepsilon$ &    &    & DNF    &  &  & \\ \hline
$\kappa$ &    &    & DNF    & DNF &  & \\ \hline
Min cluster size &    &    &     & DNF &  & \\ \hline
$\nu$ &    &    &     &  & $0.1$ & \\ \hline
kernel &    &    &     &  & RBF & \\ \hline
$f\_max$ &    &    &     &  &  &$1.0$ \\ \hline
$s\_max$ &    &    &     &  &  &$0.5$ \\ \hline
$n\_est$ &    &    &     &  &  &$50$ \\ \hline
Anomaly Score &    &    &     &  &  &$0.46$ \\  
\hline
        \end{tabular}
        }
        \caption{Parameters and results on both datasets.}
        \label{tab:Res-table-dat1_2}
                \begin{tablenotes}
            \small
            \centering
            \item[1](\textit{DNF}: \textit{Did Not Finish}.)
            \item[2] (\textit{RDF} : \textit{Radial Basis Function}.)
        \end{tablenotes}
\end{table}

\subsubsection{Autoencoder and Variational Autoencoder}
\label{AE-settings}
\begin{itemize}
\item Ensamble learing with three models for autoencoders and variational autoencoders each (see full model settings in Table \ref{tab:AutoEncoder_table_architect} for AEs and in Table \ref{tab:VAutoEncoder_table_architect} for VAEs)
\item Several architectures were tried out $\Rightarrow$ choice for the models with best looking results and lowest running time
\begin{table}[!h]
	\centering
	\resizebox{\textwidth}{!}{
	\begin{tabular}{|l|c|c|c|c|c|c|}
	\hline
		 & \multicolumn{3}{c|}{\textbf{Dataset~1}} & \multicolumn{3}{c|}{\textbf{Dataset~2}}                                               \\ \hline
		& \textbf{\begin{tabular}[c]{@{}c@{}}Model 1\end{tabular}} & \textbf{Model 2} & 	\textbf{Model 3} & \textbf{\begin{tabular}[c]{@{}c@{}}Model 1\end{tabular}} & \textbf{Model 2} & 	\textbf{Model 3} \\
	\hline 		         
	\textbf{Input layer}  & $12$ & $12$ & $12$ & $51$ & $51$ & $51$\\ \hline
\textbf{Hidden layer 1} & $6$ & $64$ & $32$ & $25$ & $64$ & $64$ \\ \hline
\textbf{Hidden layer 2}  & $3$ & $32$ & $16$ & $12$ & $32$ & $32$\\ \hline
\textbf{Latent dimension} &  $1$ & $16$ & $6$ & $6$ & $16$ & $10$ \\ \hline
\textbf{Learning rate} & $1e^{-3}$ & $1e^{-3}$ & $1e^{-3}$ & $1e^{-3}$ & $1e^{-3}$ & $1e^{-3}$ \\ \hline
\textbf{Epochs trained} & $549$ & $347$ & $450$ & $450$ & $450$ & $350$ \\
\hline
		\end{tabular}
		}
		\caption{Autoencoder architecture table for Dataset~1 and Dataset~2.}
		\label{tab:AutoEncoder_table_architect}
\end{table}

\begin{table}[!h]
	\centering
	\resizebox{\textwidth}{!}{
	\begin{tabular}{|l|c|c|c|c|c|c|}
	\hline
		& \multicolumn{3}{c|}{\textbf{Dataset~1}} & \multicolumn{3}{c|}{\textbf{Dataset~2}}                                               \\ \hline
		& \textbf{\begin{tabular}[c]{@{}c@{}}Model 1\end{tabular}} & \textbf{Model 2} & 	\textbf{Model 3} & \textbf{\begin{tabular}[c]{@{}c@{}}Model 1\end{tabular}} & \textbf{Model 2} & 	\textbf{Model 3} \\
	\hline 			         
	\textbf{Input layer}  & $12$ & $12$ & $12$ & $51$ & $51$ & $51$\\ \hline 	
\textbf{Hidden layer 1} & $6$ & $128$ & $32$ & $25$ & $128$ & $64$ \\ \hline 	
\textbf{Hidden layer 2}  & $3$ & $64$ & $16$ & $12$ & $64$ & $32$\\ \hline 	
\textbf{Latent dimension} &  $1$ & $32$ & $16$ & $6$ & $32$ & $10$ \\ \hline 	
\textbf{Learning rate} & $1e^{-3}$ & $1e^{-3}$ & $1e^{-3}$ & $1e^{-2}$ & $1e^{-2}$ & $1e^{-2}$ \\ \hline 	
\textbf{Epochs trained} & $523$ & $552$ & $537$ & $500$ & $939$ & $950$ \\
\hline
		\end{tabular}
		}
		\caption{Variational Autoencoder architecture table for Dataset~1 and Dataset~2.}
		\label{tab:VAutoEncoder_table_architect}
\end{table}

\item Table \ref{tab:AutoEncoder_table} is a summary of all the results for autoencoders and variational autoencoders on Dataset~1 and Dataset~2, respectively
\begin{table}[!h]
\centering
	\resizebox{\textwidth}{!}{
	\begin{tabular}{|l|c|c|c|c|c|c|}
	\hline
		 & \multicolumn{2}{c|}{\textbf{Autoencoder}} & \multicolumn{2}{c|}{\textbf{Variational Autoencoder}}                                               \\ \hline
		& \textbf{\begin{tabular}[c]{@{}c@{}}Dataset~1\end{tabular}} & \textbf{Dataset~2} & \textbf{\begin{tabular}[c]{@{}c@{}}Dataset~1\end{tabular}} & \textbf{Dataset~2}  \\
	\hline 	
	\textbf{Total models}  & $3$ & $3$ & $3$ & $3$ \\ \hline 	
\textbf{Threshold} & $0.5$ & $0.5$ & $0.5$ & $0.7$ \\ \hline 	
\textbf{Total anomalies}  & $224$ & $705$ & $512$ & $548$ \\ \hline 	
\textbf{Anomalies out of 4} &  $4$ & $4$ & $4$ & $4$ \\ \hline 	
\textbf{Total running time} & $2:22$min & $6$min & $4:35$min & $20$min \\
\hline
		\end{tabular}
		}
		\caption{Autoencoder and Variational Autoencoder analysis result table.}
		\label{tab:AutoEncoder_table}
\end{table}

\end{itemize}

\end{appendix}
\pagebreak
\newpage

\bibliography{Paper_Insurance_Issues_Zeldin}

\end{document}